\documentclass[reprint,amsmath,amssymb,aps,]{revtex4-2}

\usepackage{upgreek}
\usepackage{graphicx}
\usepackage{dcolumn}
\usepackage{bm}

\usepackage{siunitx}
\begin{document}

\preprint{APS/123-QED}
\title{Integrated anti-electronics for positron annihilation spectroscopy}

\author{Francesco Guatieri}
 \email{francesco.guatieri@frm2.tum.de}
\affiliation{%
 Forschungs-Neutronenquelle Heinz Maier-Leibnitz (FRM II)
 Technische Universit\"at M\"unchen
 Lichtenbergstra\ss{}e 1 85748 Garching, Germany
}%

\date{\today}

\begin{abstract}

Imaging the features of a sample using Positron Annihilation Spectroscopy (PAS) is currently achieved by rastering, i.e.\ by scanning the sample surface with a sharply focused positron beam. However, a beam of arbitrary shape (sculpted beam) would allow the application of more versatile single-pixel imaging (SPI) techniques. I introduce the design of a microelectronic device employing a 2D array of Zener diodes as an active positron moderator, capable of sculpting positron beams with $\SI{6}{\micro\meter}$ resolution. The re-emitted positrons are accelerated towards the sample through a miniaturised electrostatic lens system and reaching $\SI{100}{\nano\meter}$ resolution. The fast switch-on (\SI{90}{\pico s}) and switch-off (\SI{250}{\pico s}) time of the device would enable state-of-the-art positron annihilation lifetime spectroscopy (PALS) and PAS imaging with high spatial and temporal resolution. 
\end{abstract}


\maketitle

\section{Introduction}

Slow-positron spectroscopy is at the forefront of material science, allowing, among others, the investigation of electronic structure~\cite{ACAR1,ACAR2}, fatigue in metals~\cite{EggerFatigue2002,EggerFatigue2003}, the presence of defects~\cite{Gigl2017,Keeble2021}, and the size and distribution of pores~\cite{DickmannPorosimetry}.
These applications would benefit from high-resolution three-dimensional spectroscopic reconstruction of the sample features. The depth of the analysed feature can be determined up to a few nanometers by the well-characterised dependency of the positron implantation depth on the positron energy~\cite{Dryzek1}, a resolution that is rarely achieved with other crystallographic techniques. However, the transversal resolution of is typically much worse, since it is intrinsically limited by the diameter of the positron beam. 

Since emittance is conserved in free-space particle transport~\cite{Buon1992}, focusing a positron beam on a smaller spot eventually requires interaction with a thermal reservoir and entails sacrificing beam intensity. The most efficient way of reducing the beam radius is obtained through brightness enhancement by remoderation, i.e.\ by implanting the positrons in a crystal and employing the re-emitted ones to form a new beam: this results in a reduction of about an order of magnitude of both the beam radius and intensity~\cite{Brandes1985,Uedono2008,Hugenschmidt2015,Gigl2017}. Smaller beams can be obtained through successive stages of remoderation, which requires purposedly-designed machines and cannot be easily added to existing apparatuses. The pinnacle of positron beam focusing has been obtained by the SPM apparatus which achieves sub-micrometric focusing using three stages of remoderation~\cite{DickmannSPM}.

Conventional beam focusing techniques allow PAS two-dimensional imaging via rastering, i.e.\ by scanning the sample surface with the beam.
This technique presents some non-trivial limitations. First, the imaging resolution has to be chosen a priori, since the focusing spot size is fixed by the experimental apparatus. Second, the correct settings to focus the beam to sub-micrometric scale can only be determined through knife-edge methods~\cite{Gigl2017}, which are extremely demanding in terms of beam time. Third, if the imaging reveals non-round features within the sample that warrant further PAS investigation, selectively implanting positrons into these areas requires compressing the beam sufficiently to fit within the feature, paying the attached cost in beam intensity.
Lastly, rastering in this context is used to achieve imaging using position-insensitive detectors, yet many other single-pixel imaging (SPI) methods are known. Since current positron remoderators cannot sculpt beams into arbitrary shapes, they do not allow the use of powerful SPI techniques such as measurements in the Hadamard basis or compressed sensing (CS) methods~\cite{Candes2006,Donoho2006}.

In this work I propose to forego imaging through rastering to instead probe the sample with positron beams that have been sculpted into specific shapes and then produce the image through a reconstruction algorithm. 
An advantageous choice of beam shapes is the Hadamard basis~\cite{StreeterHadamard}.
To achieve this goal we need to build an active remoderator capable of re-emitting positrons selectively from different portions of its surface, analogous to the operation of digital micromirror devices employed in cinema projectors. We could then project the resulting sculpted beam onto a sample's surface with a resolution which is only limited by the modulator's pixel size and the precision of the optics. Such a device is equivalent to a single remoderator paired with a slit with an arbitrarily variable shape; as such, it does not achieve by itself the brightness enhancement obtained by multiple remoderation stages.
In a high-resolution positron beam apparatus it would be used to replace only the last remoderator in the stack.
Finally, I contend that a reconfigurable slit with the required flexibility and resolution can currently be realized only by implementing it as a remoderator. 

In Section~\ref{sec:PMZ} the design of a microelectronic device which can be built in a silicon substrate using established microelectronic construction techniques is introduced. Simulations show that this device can act as a remoderator capable of sculpting a beam of positron at micrometeric scale. As the beam is accelerated towards the target this resolution is then improved to sub-micrometric scales;  Section~\ref{sec:PMA} shows how this device can be combined with a compact set of electrostatic optics to realise a miniaturised high-resolution PAS imager. In section~\ref{sec:Hadamard} experimental improvements in PAS imaging that would be made possible by this device are discussed. In Section~\ref{sec:switching} it is shown that the switch-on time and switch-off time of a PMZ are sufficiently fast for positron lifetime spectroscopy. Section~\ref{sec:Conclusions} prospects some outlooks for the future expansion of this technology.

\section{Zener diodes that modulate positron emission}
\label{sec:PMZ}

In the present work I focus solely on microelectronic construction in monocrystalline silicon as it is currently cheapest and most widely available.  Moreover, interactions of slow positrons with silicon substrates have been well characterized, which allows for precise numerical simulations. Other semiconductor materials which are natively good positron remoderators are known~\cite{GaNModeration} and are an interesting venue for future analyses.

Positrons implanted in silicon with a kinetic energy of a few \SI{}{\kilo eV} thermalize in the material at a depth of a few \SI{}{\micro m}. They then undergo diffusion for about \SI{200}{\nano m} before they either annihilate with the material or reach the surface and are either trapped or re-emitted as positrons or Positronium. Strong doping in silicon is expected to reduce the positron diffusion range, although no effects have been observed at concentrations below $\SI[print-unity-mantissa=false]{1e17}{cm^{-3}}$~\cite{EplusLifetimeSi_Dopants}.

If electric fields are present inside of the silicon crystal, e.g.\ due to a doping gradient, positrons will also drift in the material~\cite{Corbel}. These diffusion-drift phenomena could allow purpose-built microelectronic devices to manipulate positrons within the silicon bulk. To this end, the most useful structure I was able to identify is that of a Zener diode. A Zener diode is a short-base diode realized with a choice of doping profile and geometry that allows its depletion zone to reach one of the bases when the diode is reverse-polarized causing a sharp increase in current called a Zener breakdown.

The intensity of the electric field within the depletion zone of any diode is capable of inducing a significant positron drift. This field is always present and, by polarizing the diode, its intensity varies. Unfortunately, this change in intensity is not sufficient to allow by itself a significant modulation of the positron drift. At the same time, polarizing the diode changes also the size of the depletion zone; this phenomenon can be exploited to significantly alter the positron drift in a portion of the substrate.
I will refer to a device designed explicitly to achieve this goal as a Positron Modulating Zener (PMZ). Notably, metal-oxide-substrate structures also allow the production and modulation of strong electric fields in microelectronics; however the interfaces between materials, unavoidable in their construction, are likely to trap positrons, which is severely detrimental to the goal of re-emitting them in vacuum.

\begin{figure}[t]
    \includegraphics[width=0.5\textwidth]{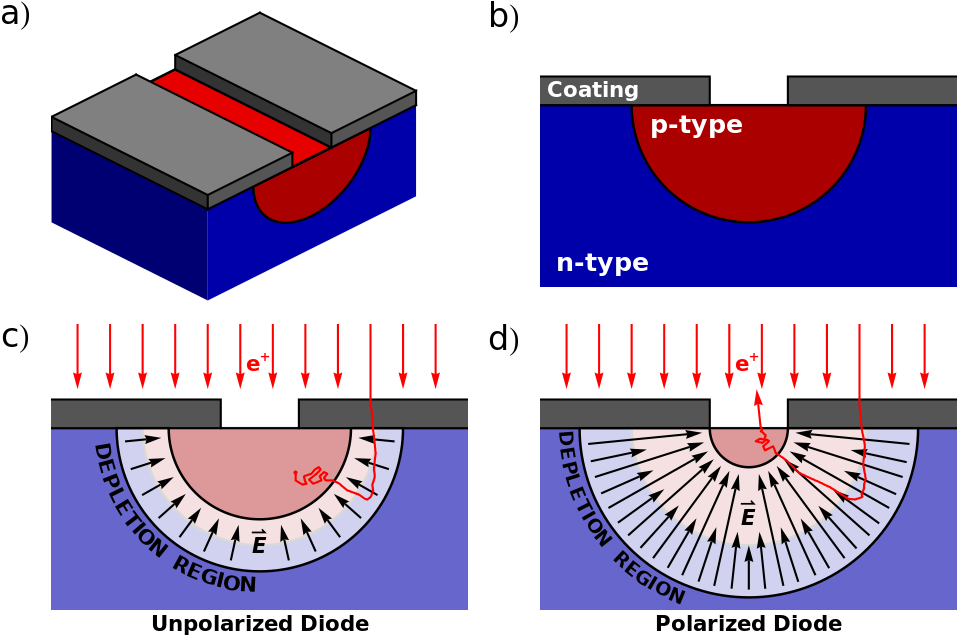}
    \caption{Schematic representations of the construction and principle of operation of a PMZ device; a) axonometric representation of the PMZ structure; b) distribution of silicon doping employed in the PMZ in transversal section; the coating material is chosen as to impede the re-emission of positrons where it is applied; c) when the diode is not polarized, most of the positrons cannot reach the surface after thermalizing; d) when the diode is reverse-polarized, the depletion zone extends further in the p-type trench, allowing its strong electric fields to guide positrons towards the gap left in the surface coating, which causes a strong increase in the re-emission probability.}
    \label{fig:Structure}
\end{figure}

Figure \ref{fig:Structure}a and b shows a schematic representation of how a PMZ could be realized. A p-type trench in the shape of a half cylinder with radius \SI{0.8}{\micro m} and a doping of $\SI{5e15}{cm^{-3}}$ is manufactured into a n-type substrate having a dopant density of $\SI{2e14}{cm^{-3}}$.
 coating is applied to the trench, made from a material (such as silicon dioxide) that would trap any positron reaching the surface, leaving only the central \SI{0.5}{\micro m} of the trench open to emit positrons. The choice of doping density ensures that the depletion zone will extend primarily in the substrate and that positron drift in the substrate is not inhibited. Figure \ref{fig:Structure}c and d show schematically the functioning mechanism of the device: when the diode is not polarized (c), the depletion region is held further away from the opening than when the diode is reverse-polarized (d). In the latter case the electric field present in the depletion zone guides the positrons closer to the aperture, letting a higher amount of them escape the device than when the diode is not polarized.

\begin{figure*}[t]
    \includegraphics[width=0.8\textwidth]{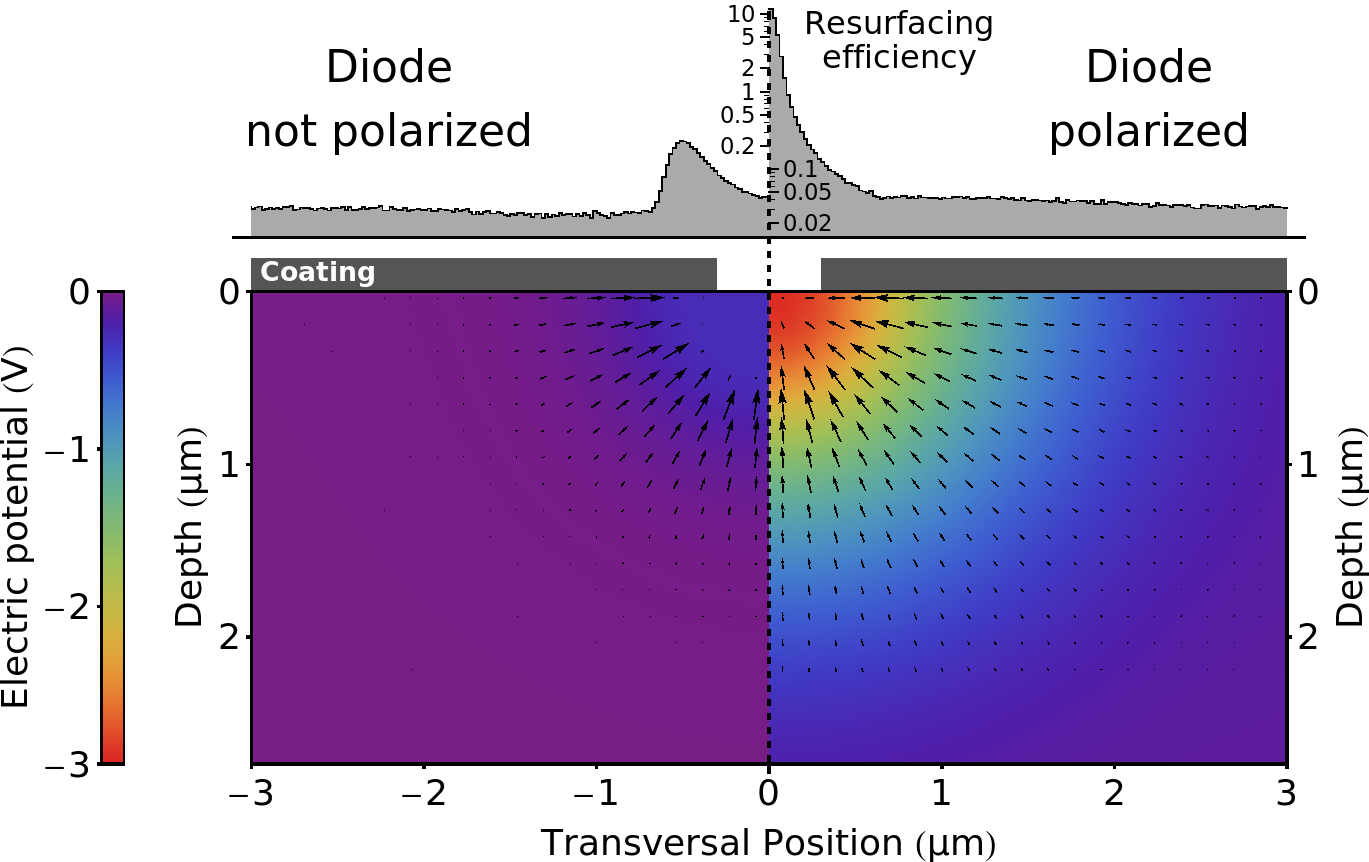}
    \caption{Finite element simulation of a PMZ response and Monte Carlo simulation of positron implantation into it. Left: PMZ not polarized. Right: PMZ reverse-polarized. The black arrows represent the direction and intensity of the electric field, color represents the electric potential. The histograms (in logarithmic scale) show the re-surfacing of positrons in the two cases; of these only positrons reaching the uncoated area at the center of the diode can be re-emitted by the PMZ.
    }
    \label{fig:PMZ}
\end{figure*}

Figure \ref{fig:PMZ} shows a finite-element simulation of the device when non polarized (left) and when polarized in reverse to the Zener breakdown voltage of \SI{2.6}{V} (right); in this regime the electric field in the region surrounding the trench has an intensity of about \SI{1}{\mega V \per m}. In these simulations a \SI{0.2}{\micro m} sigmoid-shaped doping gradient between the p and n silicon has been employed to mimic more closely the structure of a real device.

We can predict precisely the behaviour of slow positrons implanted into the device. Since the material is uniform and well characterized, the implantation of positrons can be effectively approximated by the Makhov model~\cite{Dryzek1} while post-thermalization physics can be simulated using a Monte Carlo technique. The Monte Carlo simulation has to take into account three different phenomena: drift, diffusion and annihilation. We model all three as a series of discrete interactions ~\cite{Puska1} after each of which the positrons are assigned a velocity drawn from a thermal distribution independently of their velocity prior to the interaction. Positrons travel between interactions along a parabolic trajectory, due to the external electric field, for a time $t$ randomly drawn from a exponential distribution with characteristic time $\tau_{I}$. Between interactions, positrons have a chance of annihilating given by $p_\text{ann} = \tau_I / (\tau_I+\tau_\text{ann})$, where $\tau_\text{ann}$ is the lifetime of positrons in the material.
We can verify the correctness of the simulation parameters by using it to predict positron diffusion in the material and the drift distance in presence of a constant electric field. We used $\tau_{I} = \SI{0.06}{\pico s}$ and $\tau_\text{ann} = \SI{260}{\pico s}$~\cite{EplusLifetimeSi_Dopants}, which predict a positron diffusion of $\SI{218}{\nano m}$ and a drift of $\SI{2.76}{\micro m}$ in a \SI{1}{\mega V \per m} electric field, which are in line with the values present in literature~\cite{PositronDiffusionSi,Corbel}.

The histograms placed over the device surface in Figure~\ref{fig:PMZ} show the amount and location of positrons reaching the surface of the silicon die after being implanted in the material with a kinetic energy of \SI{10}{\kilo eV}. Of the positrons implanted in a \SI{12}{\micro m}-wide region centered on the PMZ, 0.19\% 
emerge within the uncoated region when the diode is not polarized and this number increases to 13.4\%
when the diode is reverse polarized to the breakdown voltage. This allows the modulation of the positron re-emission with a contrast level of 70:1.

All the PMZ construction parameters employed here were obtained through a non-exhaustive search within the considered parameter space. It is likely that designs with better performances could be discovered through more comprehensive searches or by considering more complex geometries.

It is known that bare silicon is not an excellent emitter of positrons; from the measurements of Mills~\cite{Mills3} we can infer that about 13\% of the positrons diffusing to its surface are re-emitted into vacuum. It follows that the total remoderation efficiency of a device comprising multiple PMZs spaced in a \SI{6}{\micro m} wide grid would be only 3.5\%. It might be possible to increase this efficiency by coating the surface of the device. A natural choice could be silicon carbide (SiC), whose surface emits 30\% of the positrons reaching it~\cite{JoergsenSiC}. This would give a theoretical remoderation efficiency of 8.1\%. However, this is not realistic since interfaces between materials are known to act as traps for positrons with coefficients which are difficult to predict. The actual performance of any coating will have to be experimentally determined. 

\section{A miniaturised positron beam sculpting device}
\label{sec:PMA}

As seen in Section~\ref{sec:PMZ}, a PMZ can be realized with a size smaller than \SI{6}{\micro m}; thanks to modern microelectronic construction techniques, millions of them can be printed onto a silicon die of a few $\SI{}{\milli \meter^2}$ along with the electronic circuitry needed to address and control each one individually. I will refer to such a device as a Positron Modulating Array (PMA). Control electronics to address and switch each diode individually can be synthesized as a shift-register, a grid of x-y addressed  Metal Oxide Semiconductor (MOS) transistors or with a Charge-Coupled Device (CCD)-like structure. Regardless of the architecture of choice, the control electronics will unavoidably take up a portion of the surface that would be otherwise available for the realization of PMZs, in a measure that depends on the device design and the lithographic process employed. A way to improve this metric without increasing the resolution of the fabrication is to simply realize larger PMZs by printing comb-shaped geometries to increase their surface area.

\begin{figure}[t]
    \includegraphics[width=0.48\textwidth]{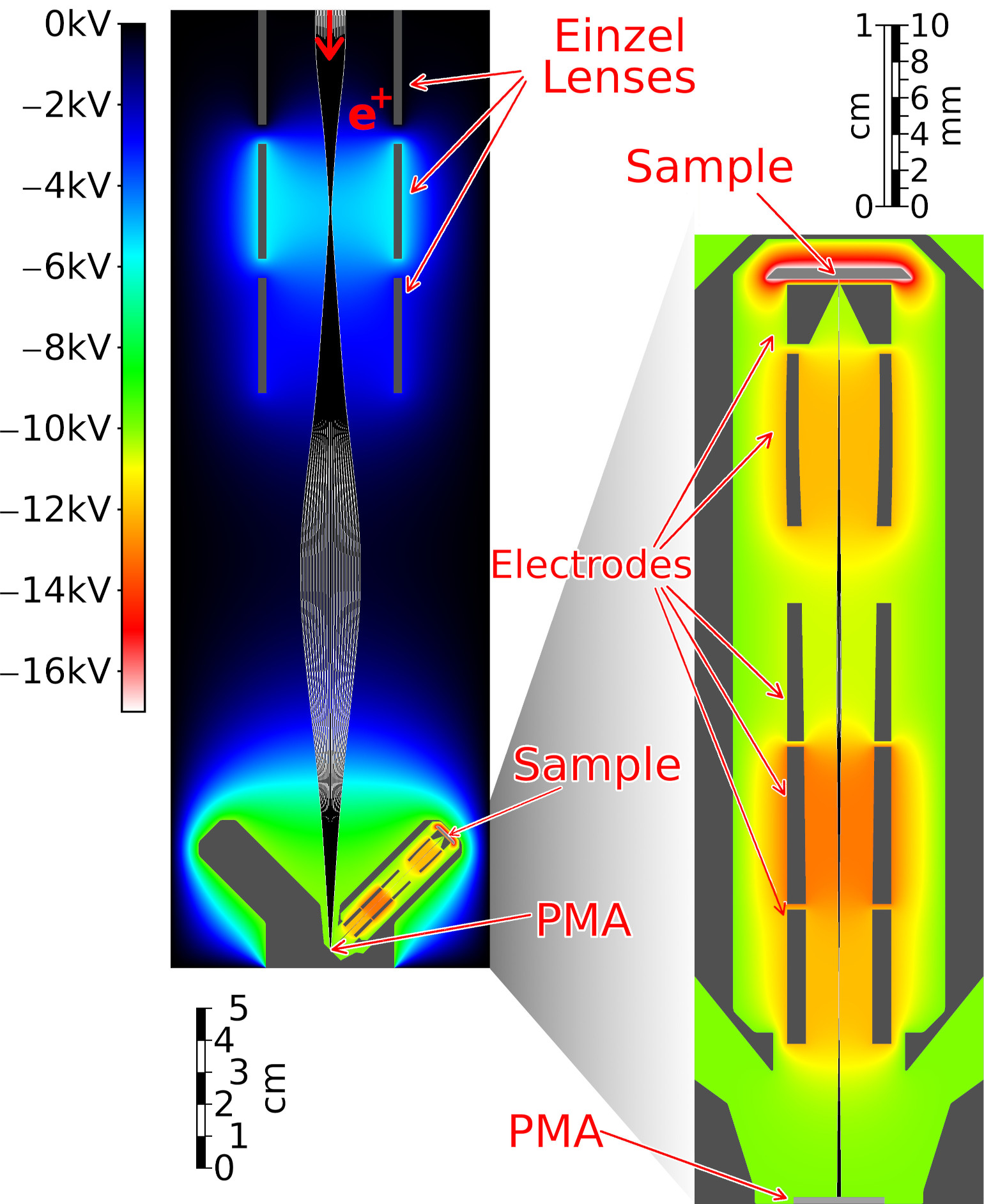}
    \caption{Possible application of a PMA device to produce a compact imaging device, as described in section~\ref{sec:PMA}. In gray, electrodes used to form the electrostatic beam optics, in color the electric potential in the free space, in black and white particle tracks from the simulated beam. Left: the primary beam is accelerated to \SI{10}{\kilo eV} and implanted into a PMA device placed at the bottom. Right: a compact beam optics designed to project and focus the positrons re-emitted from the PMA (bottom) to a sample to be imaged (top). Both the macrospic and microscopic optics are rotationally symmetric, along different axes. They have been designed so that the electric field in the volume in front of the PMA is only governed by the microscopic optics. This allows employing separate axisymmetric simulations for the two optics, that are then merged together to describe the whole apparatus.}
    \label{fig:PMA}
\end{figure}

Figure \ref{fig:PMA} shows a possible configuration in which a PMA could be installed to perform high-resolution imaging of a sample surface.
The setup focuses a \SI{10}{\kilo eV} positron beam onto the PMA using three cylindrical electrodes in a standard einzel lens configuration (left). The PMA is then used as a remoderator and to sculpt the beam; slow positrons emitted from its surface are accelerated and focused by a custom five-electrode lens system onto a sample placed \SI{5}{\centi m} away from the PMA (right).
The design shown in Figure \ref{fig:PMA} was simulated using a purpose written code to assess its expected performance in a real-world scenario, namely the CDBS apparatus installed on the NEPOMUC positron beam~\cite{Gigl2017} operated without the use of its nickel foil remoderator. Employing a PMZ spacing of \SI{6}{\micro\meter} and a \SI{7}{keV} acceleration potential a resolution of \SI{100}{\nano m} can be reached with a imaging area of $0.1 \times \SI{0.1}{\milli m}$ ($10^6$ pixels). The optics in both the macroscopic (left) and microscopic (right) portion of the machine are completely electrostatic and realized with rotationally symmetrical electrodes whose shapes have been optimized through procedural design, i.e.\ by letting an algorithm alter them based on simulation results to progress towards more and more efficient designs.

Given that imaging through a PMA does not entail rastering, the inclusion of deflector coils in the microscopic optics is not mandatory and their addition is only contingent on the necessity of correcting the beam trajectory being expected. Moreover the active nature of the PMA allows the correction of spherical aberrations at the source, within the resolution of the PMA; this requires characterizing the aberrations, which can be performed by using a CMOS detector to directly measure the secondary positron beam, as shown in a previous work~\cite{Berghold2023}. 

The small size of the PMA allows the construction of a very compact apparatus to produce, extract and focus the sculpted microbeam; the advantage of this compactness is threefold. Firstly, the small size makes it easy to keep the sample and PMA rigidly positioned with respect to each other, even in the presence of vibrations in the apparatus. Secondly, a shorter beam path reduces the distortions induced on the beam by stray electromagnetic fields and makes them easier to shield. Finally, these microbeam optics are small enough to be installed in most existing apparatuses with minimal modifications. The main challenge in realizing the apparatus presented here is the high voltage gradient present between the last electrode of the microscopic optics and the sample; this gradient is at the limit of feasibility, and as such it may be necessary to reduce it to allow for manufacturing tolerances and imperfect vacuum conditions.

\section{Single-pixel imaging techniques using a postron-modulating device}
\label{sec:Hadamard}

The use of PMA-based imaging enables the use of several experimental techniques which are inaccessible through rastering.

\begin{figure}[t]
    \includegraphics[width=0.48\textwidth]{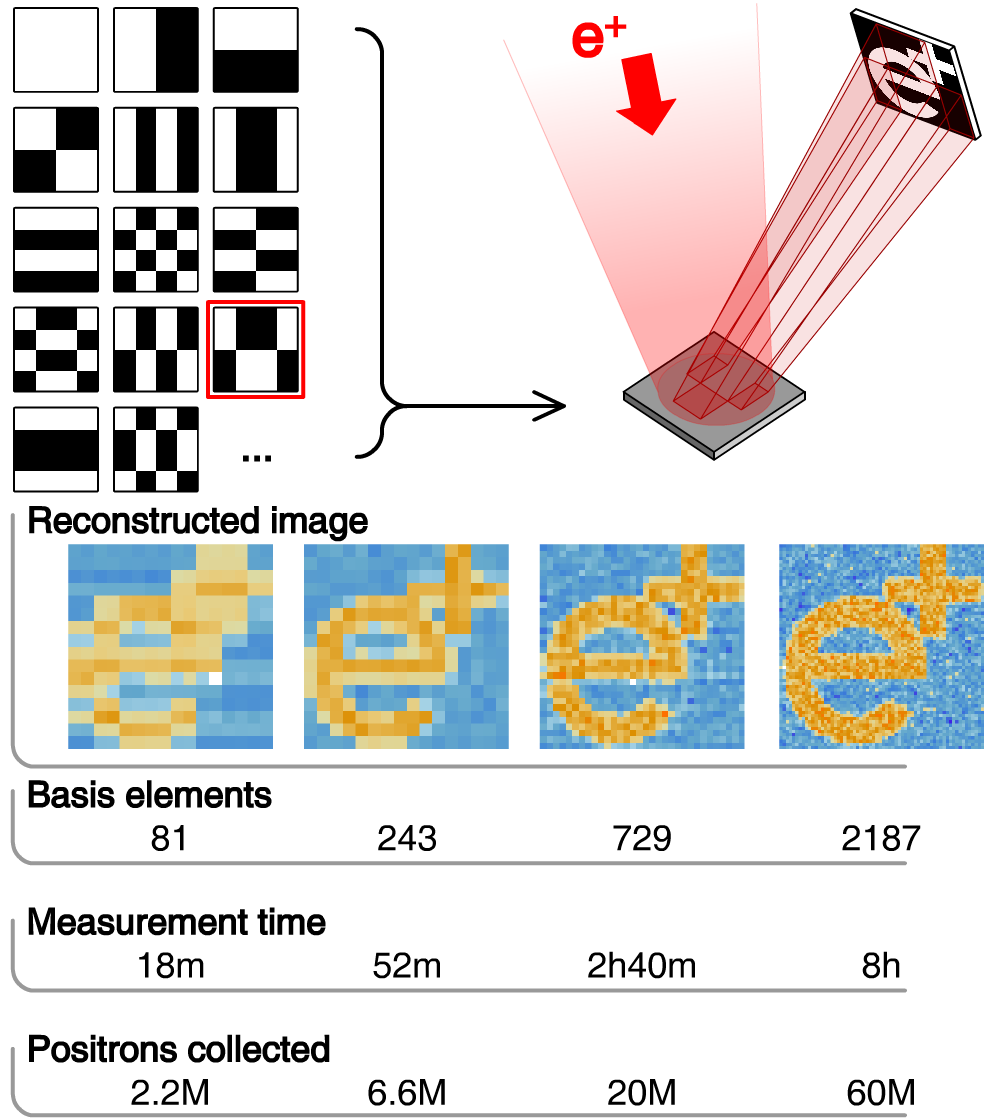}
    \caption{Above: elements of the Hadamard basis (left) are used to sculpt the beam before projecting it onto the sample (schematic representation, right). Below: a simulated experiment in which sample imaging is reconstructed at different points of the same measurement; as the measurement time increases so does the resolution of the reconstructed image, from which smaller and smaller details are progressively revealed.}
    \label{fig:Hadamard}
\end{figure}

\noindent \textbf{Progressively increasing resolution:} It is possible to use a PMA to perform SPI-like spectroscopy by projecting specific patterns onto the target (see Figure~\ref{fig:Hadamard}). The Hadamard basis is particularly convenient, as it affords discrete pixel modulations (having only $\pm 1$ entries) and, being an orthogonal transformation, does not amplify the noise under $\ell^2$ norm. Therefore, in counting experiments, as all PAS techniques ultimately are, this technique does not entail any loss of precision in the assessment of the brightness of each pixel in the reconstructed image~\cite{Wuttig2005,StreeterHadamard}. Since the elements of the Hadamard basis are composed of positive and negative sign elements, each element will require two measurements, one in which the positive-signed portion is projected onto the sample, the other in which the negative-signed portion is projected.
The greatest advantage of imaging through the Hadamard basis is that the imaging resolution increases as the measurement proceeds. This would allow an experimenter to decide to stop a measurement which has already revealed the necessary features at a lower resolution and to proceed to higher resolutions only when needed. Figure~\ref{fig:Hadamard} (bottom) shows a simulation depicting this progressive increase in resolution using a test image. I have used as reference the beam of the Coincidence Doppler-Broadening Spectroscopy (CDBS) apparatus of NEPOMUC (as in Figure~\ref{fig:PMA}), without the use of the nickel remoderator, assumed that no coating was applied to the PMA and that each element of the basis was measured for \SI{13}{s} as in~\cite{Gigl2017}. The basis element measurements have been modelled as Poisson draws.
In the four insets it is possible to see how the feature of the test target emerge with greater and greater resolution as more measurement time is spent to execute the experiment. 

\noindent \textbf{Faster beam setup:} Focusing a microbeam requires the measurement of the beam size at the target, which requires the extremely time consuming knife-edge technique~\cite{Gigl2017}. This often results in the expenditure of comparable beam time in the setup of the apparatus and the actual measurement. While we have recently proposed a possible solution based on direct imaging with CMOS sensors~\cite{Berghold2023}, it unfortunately does not extend to the submicrometer scale. The knife edge technique is ultimately a counting experiment, and as such the speed at which it can achieve a given precision is determined by the intensity of the beam employed. As the beam is reduced in diameter a cost in intensity needs to be paid regardless of the technique adopted to compress it. The possibility of sculpting the beam into arbitrary shapes allows the use of grids in place of a single knife edge, thus exploiting the entire sample surface to determine the beam focusing, which would be orders of magnitude faster than a traditional knife edge. In addition, a micrometric grid printed onto a sensor similar to those in~\cite{Berghold2023} would allow to test in parallel the alignment and focusing of the beam at different points of the desired target surface by observing the formation of Moiré patterns.

\noindent \textbf{Feature isolation:} Modern PAS techniques are increasingly moving towards \textit{in operando} measurements, that is, measurements in which, during the course of a PAS measurement, a sample is manipulated mechanically, thermally, chemically or through the exposure to light or slowly varying electromagnetic fields. \textit{In operando} measurements would benefit from the possibility of selectively studying how a feature localized in the sample responds to the external manipulation. The use of a PMA would allow preliminary imaging of the sample aimed at locating the feature of interest (e.g.\ a crack in a solid, an incorporation of a different material, a microelectronic structure). After the exact shape and location of the feature had been determined, the beam could be sculpted to implant positrons selectively into the desired feature during the \textit{in operando} measurement, thus reaching signal-to-noise ratios traditionally inaccessible. This is of particular relevance whenever it is not known whether the rest of the sample would also respond to the external manipulation, thus hindering background subtraction. Particularly amenable to this technique are man-made microstuructures, as knowledge of their expected shape makes it possible to align the sculpted beam without the need of performing a complete preliminary imaging of the sample. As an example, the study of the dielectric materials employed in MOS Field-Effect Transistors through PAS~\cite{Uedono2007} could be performed, with this technique, \textit{in operando} over the entirety of a HexFET structure. Other particularly good applications would be the study of microelectronic and micromechanic structures which can be actuated without inducing macroscopic mechanical stresses which could misalign the sample with respect to the PMA.

\noindent \textbf{Compressed sensing:} A great prominence has been gained in recent times by compressed sensing (CS) techniques, after the development of noise resilient compressed sensing algorithms~\cite{Candes2006}. Compressed sensing is the name of a family of reconstruction techniques that rely on the measurement of a low-entropy signal performed by masking it using numerous random patterns, to then reconstruct its features, under the sole assumption of them being sparse in some measurement basis~\cite{Donoho2006}, which is always the case in practical application of PAS imaging. Unfortunately, current CS techniques are not resilient against Poissonian noise~\cite{CS_Poisson}. I have tested the entire set of CS algorithms implemented in the KL1P library~\cite{KL1P} in a variety of scenarios using simulated counting experiments, and none of them have shown any significant advantage compared with the use of just the Hadamard basis.
The field is nonetheless rapidly evolving and techniques capable of working effectively with Poisson noise are under development~\cite{PoissonianCompressedSensing}, including machine learning based approaches~\cite{Niu2018}: the application of these techniques to PAS imaging is currently only possible through the use of a PMA.

\section{Switching speed of the positron-modulating Zener}
\label{sec:switching}

Among the PAS techniques, one of the most powerful is positron annihilation lifetime spectroscopy (PALS), which works by implanting a pulsed beam of positrons into a sample to then measure the time taken by the implanted positrons to annihilate. In this application, the length and sharpness of the positron pulses combined with the jitter of the detectors determines the resolution function. Ideally, the width of the resolution function should be kept to the order of \SI{100}{\pico s}~\cite{Saito100ps}.
As we shall see, the expected transient time of a PMA is fast enough for it to be used either to sharpen the output of a positron buncher/chopper, or to replace it altogether.

\begin{figure}[t]
\includegraphics[width=0.5\textwidth]{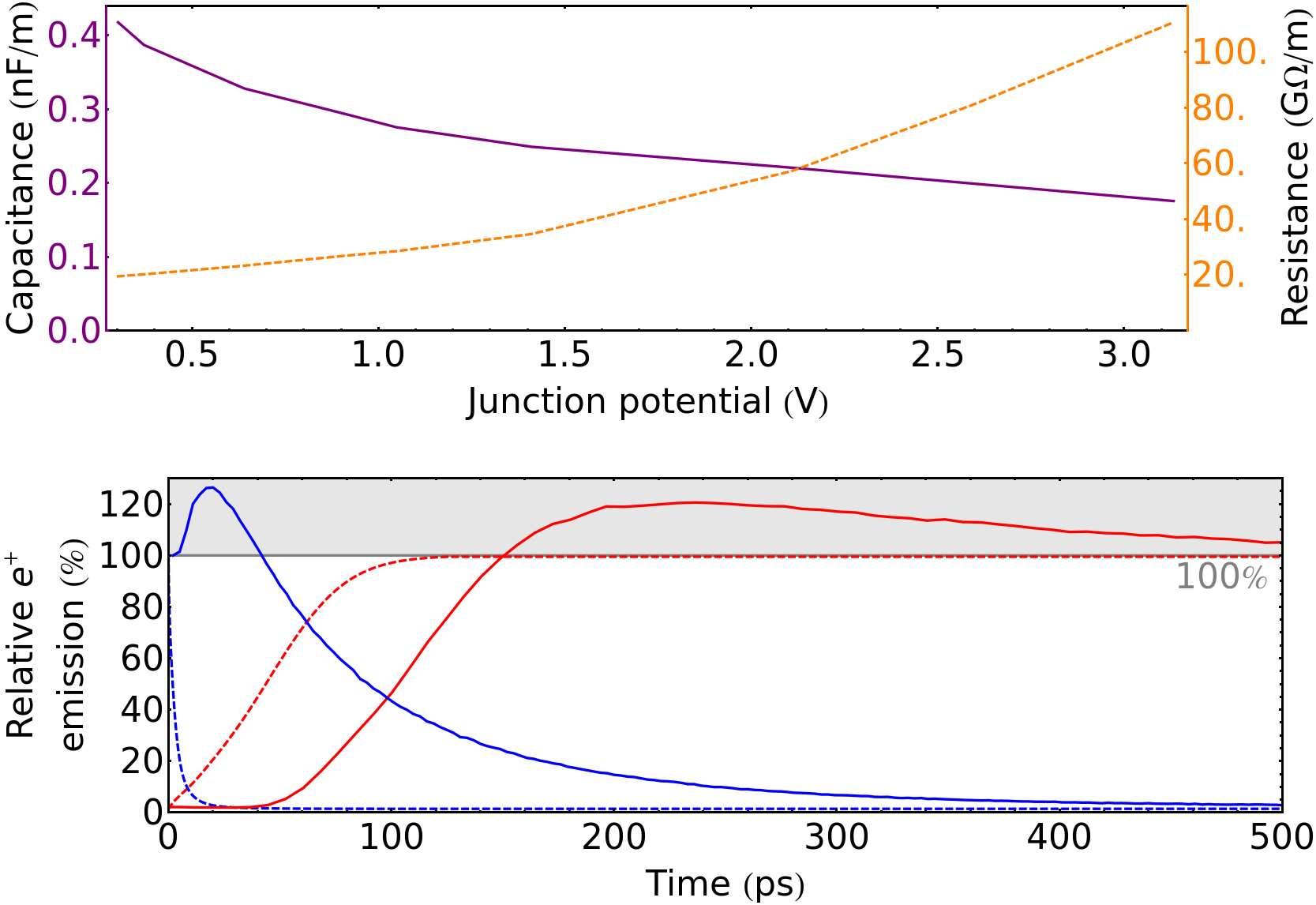}
\caption{\label{fig:Transient} Top: expected resistance (orange) and capacitance (purple) per unit length of the p-type trench. Bottom: simulated switch-on (red) and switch-off (blue) transients of the actuation of a \SI{5}{\micro m}-long PMZ contacted on both ends, the dashed line indicates the actuation as expected purely from an electronic standpoint. The solid lines are the result of a full simulation which takes into account the drift of positrons in the diode. The full simulation exhibits both a much longer switch-off transient than the purely electronic simulation and overshoots of the steady state emission in both transients.}
\end{figure}

The electronic switching time of a PMZ depends on how quickly the charge needed to polarize or de-polarize it can be transported along the p-type channel in the center of the trench. Both the capacitance and resistance of the channel depend on the local polarization status of the diode. These can be computed from the number of free carriers predicted by the finite element simulation employed in section \ref{sec:PMZ}. The resulting curves are shown in figure \ref{fig:Transient} (top). To realize a fast-switching PMZ we want to limit its length, so for this simulations a \SI{5}{\micro m} long PMZ contacted on both sides was considered. In general we can expect the electronic switching time to scale approximately with the square of the diode length. An exact solution of the polarization/depolarization transient would require a time-dependent three dimensional finite element simulation. However, in this context it is practical to approximate it by using the voltage-dependent resistance and capacitance of the trench to write a diffusion equation which can then be solved numerically. The result is a wave of polarization/depolarization which travels along the length of the diode with decreasing speed; the electronic transient in the case of a \SI{5}{\micro m}-long PMZ is shown with dashed lines Figure~\ref{fig:Transient} (bottom), with the local diode polarization translated into a positron re-emission by using the steady-state emission of the diode at different polarization. If computed this way the switch-on time (time to reach 95\% of the positron emission) is \SI{90}{\pico s} while the much faster switch-off time (time to reach 5\% of the emissions) is \SI{13}{\pico s}. The difference in time is due to the fact that, as the diode polarizes, the channel gets pinched by the diode depletion zone, increasing the resistance of the channel and making it more difficult to carry the charge needed to polarize the diode further along its length. This model does not, unfortunately, capture all the relevant physics of the PMZ transient, since the time the positrons need to diffuse in the material is neglected.

To simulate the diffusion process we need to run a Monte Carlo simulation in the three-dimensional volume occupied by the PMZ using a time-dependent electric field and a flux of implanted positrons uniformly distributed within the simulation time window, which extends to the nanosecond prior to the switching time to allow for the PMZ to reach a steady state before switching. The result of this simulation is shown with solid lines in Figure~\ref{fig:Transient} (bottom). When employing this simulation technique the switch-on time is still \SI{90}{\pico s}, albeit with an added \SI{60}{\pico s} delay, while the switch-off time increases to \SI{250}{\pico s}. The most striking feature is the presence of overshoots in both transients. The overshoot in the switch-on transient is due to the accumulation of positrons in the silicon bulk while the PMZ is not polarized. During the steady state with the diode polarized, positrons continuously drift towards the surface and are re-emitted with a rate that is dictated by how quickly new positrons are implanted in the material. This reduces the permanence time of positrons in silicon which would otherwise be (with the exception of positrons implanted close to the surface) given by the positron lifetime in the solid. As a consequence, the number of positrons present in the silicon bulk at any given time is higher when the diode is not polarized. Shortly after polarization this excess positrons are flushed out, thus briefly overshooting the steady state emission rate. The overshoot at switch-off is due to a different phenomenon: the electric field within a polarized PMZ is mostly directed towards the surface (see Figure~\ref{fig:PMZ}d) except in extreme proximity to the surface where it changes direction and acquires a small component pointing away from the material surface. This small component effectively hinders the re-emission of positrons, even though most will be able to overcome it and exit the material. When the diode polarization is removed, this hindrance is removed too and thus positrons that were already close to the surface will more easily escape the silicon, once again exceeding briefly the steady state emission rate.

All considered, the expected transient time of the PMZ make it a good candidate to act by itself as a chopper/buncher in a compact PALS apparatus or to work in combination with an external chopper/buncher to increase its time resolution.

\section{Conclusions and outlook}
\label{sec:Conclusions}

I have presented the design of an active remoderator capable of modulating the re-emission of positrons from any point of its surface, sculpting a positron beam into an arbitrary shape. Numerical simulations show that a resolution of \SI{6}{\micro m} can be achieved with a contrast of 70:1. Individual emitters of the device can be switched on in \SI{90}{\pico s} and switched off in \SI{250}{\pico s}. I have then presented a possible design of a microbeam PAS imaging optic making use of this device. Simulations show that when positrons are implanted into the target with an energy of \SI{7}{\kilo eV} an imaging resolution of \SI{100}{\nano m} can be achieved over an imaging area of $0.1 \times \SI{0.1}{\milli m^2}$. This device is extremely compact, with a distance of \SI{5}{\centi m} between the PMA and the sample, which at the same time makes it more easily shielded from ambient noise and compatible with existing apparatuses.

The use of this device would enable several new experimental techniques. Firstly, the possibility of selecting the desired imaging resolution as the measurement is performed, based on the measurement results. Secondly a faster calibration of the beam optics, in particular when sub-micrometer resolutions are to be achieved. Thirdly, the isolation of non-round features which can then be measured with high signal-to-noise ratios using PAS techniques. Finally, it enables the use of compressed sensing with PAS techniques.

The modern microelectronic industry has produced a wealth of different integrated devices, with a rich taxonomy of structures and functions. It is my belief that many devices might exist within this assortment, with potential applications to low-energy positron techniques performed through the manipulation of positrons in a material bulk. These applications could span from the construction of positron-to-positronium converters whose positronium production can be spatially and temporally tuned, to the study of samples embedded directly into the silicon bulk, to the formation of a positronium Bose-Einstein condensate~\cite{MillsBEC}. Two main obstacles hinder the exploration of these opportunities: the difficulty of simulating these scenarios and the high costs attached to the custom production of these devices. It is my hope that the present work will spark some interest in trying to overcome both. In the meantime, since the use of PMAs prescinds from the specific PAS techniques employed, the identification of their feasibility is a potential benefit to the whole of known positron spectroscopy techniques.

\begin{acknowledgments}

Firstly it is imperative that I give my thanks to Dr. Davide Orsucci, for his relentless help in reviewing the drafts of my papers, the invaluable insights that without fail he provides me with and his unwavering patience in tolerating my pedantry. On top of that I would like to thank Prof. Christoph Hugenschmidt, Dr. Daniel Russell and Dr. Alessandro Moia for the advice they provided during the preparation of this paper.

\end{acknowledgments}

\bibliography{Antielectronics}

\end{document}